\documentclass[12pt, a4paper]{article}
\usepackage{geometry, amsmath,amssymb,amsthm,amsxtra,overpic,bbm,bm,epsfig,sub
figure}
\usepackage{mathrsfs}
\usepackage{graphicx}
\usepackage{color}
\usepackage{comment}
\usepackage{epstopdf}
\usepackage{float}

\usepackage{slashed,stmaryrd}

\usepackage{multirow}

\def\thefootnote{\fnsymbol{footnote}}

\begin{document}

\vspace{0.2cm}

\begin{center}
{\large\bf A new Wolfenstein-like expansion of lepton flavor mixing \\
towards understanding its fine structure}
\end{center}

\vspace{0.2cm}

\begin{center}
{\bf Zhi-zhong Xing$^{1,2}$}
\footnote{E-mail: xingzz@ihep.ac.cn}
\\
{\small $^{1}$Institute of High Energy Physics and School of Physical Sciences, \\
University of Chinese Academy of Sciences, Beijing 100049, China \\
$^{2}$Center of High Energy Physics, Peking University, Beijing 100871, China}
\end{center}

\vspace{2cm}
\begin{abstract}
Taking the tri-bimaximal flavor mixing pattern as a particular basis,
we propose a new way to expand the $3\times 3$ unitary
Pontecorvo-Maki-Nakagawa-Sakata (PMNS) lepton flavor mixing matrix $U$
in powers of the magnitude of its smallest element
$\xi \equiv \left|U^{}_{e 3}\right| \simeq 0.149$.
Such a Wolfenstein-like parametrization of $U$ allows us to easily describe
the salient features and fine structures of flavor mixing and CP violation,
both in vacuum and in matter.
\end{abstract}

\def\thefootnote{\arabic{footnote}}
\setcounter{footnote}{0}

\newpage

\section{Motivation}

Among the three Euler-like rotation angles of the
Cabibbo-Kobayashi-Maskawa (CKM) quark flavor mixing matrix
$V$~\cite{Cabibbo:1963yz,Kobayashi:1973fv},
it is the largest one --- the Cabibbo angle $\theta^{}_{\rm C}
\simeq 13^\circ$ that was most accurately measured from the very
beginning~\cite{ParticleDataGroup:2022pth}. That is why Wolfenstein
proposed a remarkable parametrization of the CKM matrix $V$ in 1983
by expanding its nine matrix elements in powers of a small parameter
$\lambda \equiv \sin\theta^{}_{\rm C} \simeq 0.225$~\cite{Wolfenstein:1983yz},
from which the hierarchical structure of quark flavor mixing can be well
understood. For example, one may easily arrive at the four-layered ordering
\begin{eqnarray}
\underline{|V^{}_{tb}| > |V^{}_{ud}| > |V^{}_{cs}|} \gg
\underline{|V^{}_{us}| > |V^{}_{cd}|} \gg
\underline{|V^{}_{cb}| > |V^{}_{ts}|} \gg
\underline{|V^{}_{td}| > |V^{}_{ub}|}
\end{eqnarray}
of respective ${\cal O}(1)$, ${\cal O}(\lambda)$,
${\cal O}(\lambda^2)$ and ${\cal O}(\lambda^3)$ as a natural consequence
of the unitarity of $V$~\cite{Xing:1996it}. In particular, the CKM
matrix $V$ approaches the unique identity matrix $I$
in the $\lambda \to 0$ limit, implying that the up- and down-type quark
sectors should have an underlying parallelism between their flavor textures.
Such a conceptually interesting limit is quite suggestive, and it has
widely been considered for explicit model building~\cite{Xing:2020ijf}.

In comparison, two of the three Euler-like rotation angles of the
Pontecorvo-Maki-Nakagawa-Sakata (PMNS) lepton flavor
mixing matrix $U$ (i.e., $\theta^{}_{12} \simeq 33.4^\circ$ and
$\theta^{}_{23} \sim 45^\circ$~\cite{ParticleDataGroup:2022pth})
are so large that a naive Wolfenstein-like parametrization of $U$ seems
quite unlikely
\footnote{But see, e.g., Refs.~\cite{Xing:2002az,Kaus:2002zm,Gupta:2013vva},
for the early attempts in this regard.}.
A very real possibility is that the leading-order term of the PMNS
matrix $U$ is a constant flavor mixing pattern $U^{}_0$ consisting of two
large angles and originating from a kind of discrete flavor symmetry, while
the smallest flavor mixing angle $\theta^{}_{13} \simeq 8.6^\circ$ and
CP-violating effects arise after small perturbations or quantum corrections
to $U^{}_0$ are taken into account~\cite{Fritzsch:1995dj,Fritzsch:1998xs}.
From the point of view of model
building~\cite{Xing:2020ijf,Feruglio:2019ybq,King:2017guk,Ding:2024ozt}, the
most popular choice of $U^{}_0$ has been the tri-bimaximal flavor mixing
pattern $U^{}_{\rm TBM}$~\cite{Harrison:2002er,Xing:2002sw,He:2003rm}
which predicts $\theta^{(0)}_{12} = \cot\sqrt{2} \simeq 35.26^\circ$,
$\theta^{(0)}_{13} = 0^\circ$ and $\theta^{(0)}_{23} = 45^\circ$. Given
the smallness of $\theta^{}_{13}$, several attempts have been made
along the above line of thought to expand the $U^{}_{\rm TBM}$-based
PMNS matrix $U$ in powers of the small Wolfenstein
parameter $\lambda$ that is borrowed from quark flavor mixing (see, e.g.,
Refs.~\cite{Li:2004dn,King:2007pr,King:2009qt,King:2012vj,Liu:2014gla}).

Guided by the best-fit values and $1\sigma$ intervals of three
lepton flavor mixing angles extracted from a global analysis of
the currently available experimental data on neutrino oscillations~\cite{Gonzalez-Garcia:2021dve,Capozzi:2021fjo},
\begin{eqnarray}
{\rm NMO:} \quad \theta^{}_{12} = 0.583^{+0.013}_{-0.013} \; , \quad
\theta^{}_{13} = 0.150^{+0.002}_{-0.002} \; , \quad
\theta^{}_{23} = 0.736^{+0.019}_{-0.015} \;
\end{eqnarray}
for the normal mass ordering (NMO) of three active neutrinos, or
\begin{eqnarray}
{\rm IMO:} \quad \theta^{}_{12} = 0.583^{+0.013}_{-0.013} \; , \quad
\theta^{}_{13} = 0.150^{+0.002}_{-0.002} \; , \quad
\theta^{}_{23} = 0.855^{+0.018}_{-0.020} \;
\end{eqnarray}
in the inverted mass ordering (IMO) case, we find that the smallest
angles $\theta^{}_{13}$ is most accurately determined and thus
suitable for serving as an optimal expansion parameter
\footnote{It is worth pointing out that the previous phenomenological
conjecture $|U^{}_{e 3}| = \lambda/\sqrt{2} \simeq 0.159$~\cite{Giunti:2002ye}
is no more favored by the Daya Bay precision measurements (i.e.,
$\sin\theta^{}_{13} \simeq 0.149$)~\cite{DayaBay:2018yms}.}.
So we are going to study an expansion of the $U^{}_{\rm TBM}$-based
PMNS matrix $U$ in powers of the small parameter
$\xi \equiv |U^{}_{e 3}| = \sin\theta^{}_{13} \simeq 0.149
\simeq 0.662 \lambda$. Moreover, the best-fit values of $\theta^{}_{12}$
and $\theta^{}_{23}$ lead us to
\begin{eqnarray}
{\rm NMO:} \quad  \theta^{}_{12} - \theta^{(0)}_{12} & \simeq & -0.032 \simeq -1.463
\hspace{0.04cm} \xi^2 \; ,
\nonumber \\
\theta^{}_{23} - \theta^{(0)}_{23} & \simeq & -0.049 \simeq -2.225
\hspace{0.04cm} \xi^2 \; ; \hspace{0.5cm}
\end{eqnarray}
as well as
\begin{eqnarray}
{\rm IMO:} \quad \theta^{}_{12} - \theta^{(0)}_{12} & \simeq & -0.032 \simeq -1.463
\hspace{0.04cm} \xi^2 \; ,
\nonumber \\
\theta^{}_{23} - \theta^{(0)}_{23} & \simeq & +0.070 \simeq +3.135
\hspace{0.04cm} \xi^2 \; . \hspace{0.5cm}
\end{eqnarray}
Eqs.~(4) and (5) imply
that the observed values of $\theta^{}_{12}$ and $\theta^{}_{23}$
are most likely to deviate from their respective tri-bimaximal flavor
mixing limits at the level of ${\cal O}(\xi^2)$. This observation
provides us with a new angle of view, which is quite different from
those in the previous attempts, to expand $U$ in the basis of $U^{}_{\rm TBM}$.

In what follows we shall propose a new expansion of the PMNS matrix
$U$ by starting from the standard Euler-like parametrization of $U$
and taking account of
\begin{itemize}
\item     $\sin\theta^{}_{13} \equiv \xi \simeq 0.149$ as the
lepton flavor mixing expansion parameter;

\item     $\theta^{}_{12} = \theta^{(0)}_{12} - A \xi^2$ with
$\theta^{(0)}_{12} = \cot\sqrt{2} \simeq 35.26^\circ$ and
$A \sim {\cal O}(1)$;

\item     $\theta^{}_{23} = \theta^{(0)}_{23} - B \xi^2$ with
$\theta^{(0)}_{23} = 45^\circ$ and $|B| \sim {\cal O}(1)$.
\end{itemize}
It is obvious that $B \neq 0$ characterizes the effect of $\mu$-$\tau$
permutation symmetry breaking of $U$, and the sign of $B$ determines
the octant of $\theta^{}_{23}$~\cite{Xing:2015fdg,Xing:2022uax}.
In this case the PMNS matrix $U$ will be expressed in terms of the tri-bimaximal
flavor mixing pattern $U^{}_{\rm TBM}$, the three real parameters
$\xi$, $A$ and $B$, and the poorly known CP-violating phase $\delta$.
Here we leave aside the two possible extra CP phases associated with the
Majorana nature of massive neutrinos, as they are completely unknown and
have little effect on the fine structure of $U$.

\section{The expansion of $U$}

As advocated by the Particle Data Group (PDG), the standard Euler-like
parametrization of the unitary $3\times 3$ PMNS matrix $U$ is
explicitly of the form~\cite{ParticleDataGroup:2022pth}
\begin{eqnarray}
U = \left(\begin{matrix}
c^{}_{12} c^{}_{13} & s^{}_{12} c^{}_{13} &
s^{}_{13} e^{-{\rm i} \delta} \cr
-s^{}_{12} c^{}_{23} - c^{}_{12}
s^{}_{13} s^{}_{23} e^{{\rm i} \delta} & c^{}_{12} c^{}_{23} -
s^{}_{12} s^{}_{13} s^{}_{23} e^{{\rm i} \delta} & c^{}_{13}
s^{}_{23} \cr
s^{}_{12} s^{}_{23} - c^{}_{12} s^{}_{13} c^{}_{23}
e^{{\rm i} \delta} & ~ - c^{}_{12} s^{}_{23} - s^{}_{12} s^{}_{13}
c^{}_{23} e^{{\rm i} \delta} ~ &  c^{}_{13} c^{}_{23} \cr
\end{matrix} \right) \; ,
\end{eqnarray}
in which $c^{}_{ij} \equiv \cos\theta^{}_{ij}$ and
$s^{}_{ij} \equiv \sin\theta^{}_{ij}$
(for $ij = 12, 13, 23$) with $\theta^{}_{ij}$ lying in the first quadrant,
$\delta$ denotes the irreducible CP-violating phase responsible for
CP violation in neutrino oscillations, and possible additional CP
phases of the Majorana nature can always be factored out on the right-hand
side of $U$ and thus have been omitted.
Following the strategy outlined above for a Wolfenstein-like
expansion of the $U^{}_{\rm TBM}$-based PMNS matrix $U$, we have
\begin{eqnarray}
s^{}_{13} & \equiv & \xi \; ,
\nonumber \\
c^{}_{13} & = & 1 - \frac{1}{2} \xi^2 - \frac{1}{8} \xi^4
+ {\cal O}(\xi^6) \; ,
\nonumber \\
s^{}_{12} & = & \frac{1}{\sqrt 3} \left[1 - \sqrt{2} A \xi^2
- \frac{1}{2} A^2 \xi^4\right]
+ {\cal O}(\xi^6) \; ,
\nonumber \\
c^{}_{12} & = & \frac{2}{\sqrt 6} \left[1 + \frac{1}{\sqrt{2}} A \xi^2
- \frac{1}{2} A^2 \xi^4\right]
+ {\cal O}(\xi^6) \; , \hspace{0.5cm}
\nonumber \\
s^{}_{23} & = & \frac{1}{\sqrt 2} \left[1 - B \xi^2
- \frac{1}{2} B^2 \xi^4\right]
+ {\cal O}(\xi^6) \; ,
\nonumber \\
c^{}_{23} & = & \frac{1}{\sqrt 2} \left[1 + B \xi^2
- \frac{1}{2} B^2 \xi^4\right]
+ {\cal O}(\xi^6) \; .
\end{eqnarray}
Substituting Eq.~(7) into Eq.~(6), we immediately arrive at the nine elements of
$U$ in the PDG-advocated phase convention as follows:
\begin{eqnarray}
U^{}_{e 1} & = & \frac{2}{\sqrt 6} - \frac{1 - \sqrt{2} A}{\sqrt 6} \xi^2
- \frac{1 + 2 \sqrt{2} A + 4 A^2}{4 \sqrt{6}} \xi^4 \; ,
\nonumber \\
U^{}_{e 2} & = & \frac{1}{\sqrt 3} - \frac{1 + 2\sqrt{2} A}{2 \sqrt{3}} \xi^2
- \frac{1 - 4\sqrt{2} A + 4 A^2}{8 \sqrt{3}} \xi^4 \; ,
\nonumber \\
U^{}_{e 3} & = & \xi e^{-{\rm i}\delta} \; ,
\nonumber \\
U^{}_{\mu 1} & = & -\frac{1}{\sqrt 6} - \frac{1}{\sqrt 3} \xi e^{{\rm i}\delta}
+ \frac{\sqrt{2} A - B}{\sqrt 6} \xi^2
- \frac{A - \sqrt{2} B}{\sqrt 6} \xi^3 e^{{\rm i}\delta}
+ \frac{A^2 + B^2 + 2\sqrt{2} AB}{2 \sqrt{6}} \xi^4 \hspace{0.5cm}
\nonumber \\
&& + \frac{A^2 + B^2 + \sqrt{2} AB}{2 \sqrt{3}} \xi^5 e^{{\rm i}\delta} \; ,
\nonumber \\
U^{}_{\mu 2} & = & \frac{1}{\sqrt 3} - \frac{1}{\sqrt 6} \xi e^{{\rm i}\delta}
+ \frac{A + \sqrt{2} B}{\sqrt 6} \xi^2
+ \frac{\sqrt{2} A + B}{\sqrt 6} \xi^3 e^{{\rm i}\delta}
- \frac{A^2 + B^2 - \sqrt{2} AB}{2 \sqrt{3}} \xi^4
\nonumber \\
&& + \frac{A^2 + B^2 + 2\sqrt{2} AB}{2 \sqrt{6}} \xi^5
e^{{\rm i}\delta} \; ,
\nonumber \\
U^{}_{\mu 3} & = & \frac{1}{\sqrt 2} - \frac{1 + 2 B}{2 \sqrt{2}} \xi^2
- \frac{\left(1 - 2 B\right)^2}{8 \sqrt{2}} \xi^4 \; ,
\nonumber \\
U^{}_{\tau 1} & = & \frac{1}{\sqrt 6} - \frac{1}{\sqrt 3} \xi e^{{\rm i}\delta}
- \frac{\sqrt{2} A + B}{\sqrt 6} \xi^2
- \frac{A + \sqrt{2} B}{\sqrt 6} \xi^3 e^{{\rm i}\delta}
- \frac{A^2 + B^2 - 2\sqrt{2} AB}{2 \sqrt{6}} \xi^4
\nonumber \\
&& + \frac{A^2 + B^2 - \sqrt{2} AB}{2 \sqrt{3}} \xi^5
e^{{\rm i}\delta} \; ,
\nonumber \\
U^{}_{\tau 2} & = & -\frac{1}{\sqrt 3} - \frac{1}{\sqrt 6} \xi e^{{\rm i}\delta}
- \frac{A - \sqrt{2} B}{\sqrt 6} \xi^2
+ \frac{\sqrt{2} A - B}{\sqrt 6} \xi^3 e^{{\rm i}\delta}
+ \frac{A^2 + B^2 + \sqrt{2} AB}{2 \sqrt{3}} \xi^4
\nonumber \\
&& + \frac{A^2 + B^2 + 2\sqrt{2} AB}{2 \sqrt{6}} \xi^5 e^{{\rm i}\delta} \; ,
\nonumber \\
U^{}_{\tau 3} & = & \frac{1}{\sqrt 2} - \frac{1 - 2 B}{2 \sqrt{2}} \xi^2
- \frac{\left(1 + 2 B\right)^2}{8 \sqrt{2}} \xi^4 \; ,
\end{eqnarray}
up to ${\cal O}(\xi^6)$ or equivalently ${\cal O}(10^{-5})$. This degree
of precision and accuracy for the elements of $U$ should be good
enough to confront the present and future precision measurements of
various neutrino oscillation channels. Some comments and discussions are in order.
\begin{itemize}
\item     The two off-diagonal asymmetries of the unitary PMNS matrix $U$,
which largely characterize its geometrical structure about the
$U^{}_{e 1}$-$U^{}_{\mu 2}$-$U^{}_{\tau 3}$ and
$U^{}_{e 3}$-$U^{}_{\mu 2}$-$U^{}_{\tau 1}$ axes, are given by
\begin{eqnarray}
{\cal A}^{}_1 & \equiv & |U^{}_{e 2}|^2 - |U^{}_{\mu 1}|^2 =
|U^{}_{\mu 3}|^2 - |U^{}_{\tau 2}|^2 = |U^{}_{\tau 1}|^2 - |U^{}_{e 3}|^2
\nonumber \\
& \simeq & +\frac{1}{6} - \frac{\sqrt 2}{3} \xi \cos\delta - \frac{1}{3}
\left(2 + 2 A^2 + B^2 + 2\sqrt{2} A B\right) \xi^2 \; , \hspace{0.5cm}
\nonumber \\
{\cal A}^{}_2 & \equiv & |U^{}_{e 2}|^2 - |U^{}_{\mu 3}|^2 =
|U^{}_{\mu 1}|^2 - |U^{}_{\tau 2}|^2 = |U^{}_{\tau 3}|^2 - |U^{}_{e 1}|^2
\nonumber \\
& \simeq & -\frac{1}{6} + \frac{1}{6} \left(1 - 4\sqrt{2} A
+ 6 B\right) \xi^2 \; .
\end{eqnarray}
In comparison, the corresponding off-diagonal asymmetries of the
CKM quark flavor mixing matrix $V$ are respectively of ${\cal O}(\lambda^6)$
and ${\cal O}(\lambda^2)$~\cite{Xing:1996it}. So the PMNS matrix $U$
is geometrically not so symmetrical as the CKM matrix $V$. Given the
fact that either ${\cal A}^{}_1 = 0$ or ${\cal A}^{}_2 = 0$ would
imply the congruence of three pairs of the PMNS unitarity triangles
in the complex plane~\cite{Xing:2002sx}, we find that the relatively
large off-diagonal asymmetries of $U$ means that its six unitarity
triangles are not very similar to one another in shape.

\item     The three $\mu$-$\tau$ interchange asymmetries of the PMNS
matrix $U$, which describe small effects of the $\mu$-$\tau$ flavor symmetry
breaking, are found to be
\begin{eqnarray}
\Delta^{}_1 & \equiv & |U^{}_{\tau 1}|^2 - |U^{}_{\mu 1}|^2
\simeq - \frac{2}{3} B \xi^2 \left[1 - 2\left(1 + \sqrt{2} A\right) \xi^2\right]
-\frac{2}{3} \xi \left(\sqrt{2} - A \xi^2\right) \cos\delta \; , \hspace{0.5cm}
\nonumber \\
\Delta^{}_2 & \equiv & |U^{}_{\tau 2}|^2 - |U^{}_{\mu 2}|^2
\simeq - \frac{2}{3} B \xi^2 \left[2 - \left(1 - 2\sqrt{2} A\right) \xi^2\right]
+\frac{2}{3} \xi \left(\sqrt{2} - A \xi^2\right) \cos\delta \; ,
\nonumber \\
\Delta^{}_3 & \equiv & |U^{}_{\tau 3}|^2 - |U^{}_{\mu 3}|^2
\simeq 2 B \xi^2 \left(1 - \xi^2\right) \; .
\end{eqnarray}
Of course, $\Delta^{}_1 + \Delta^{}_2 + \Delta^{}_3 = 0$ holds, as
assured by the unitarity of $U$. One can simply see that
$\Delta^{}_1 = \Delta^{}_2 = \Delta^{}_3 = 0$ requires both
$B = 0$ and $\delta = \pm \pi/2$, the conditions that allow $U$ to have the
exact $\mu$-$\tau$ reflection symmetry~\cite{Harrison:2002et}. The preliminary
T2K measurement hints at $\delta \sim -\pi/2$~\cite{T2K:2023smv}, and thus
$\Delta^{}_1$, $\Delta^{}_2$ and $\Delta^{}_3$ are all expected to be
of ${\cal O}(10^{-2})$. It is therefore expected that the $\mu$-$\tau$ permutation
or reflection symmetry may serve as a minimal lepton flavor symmetry which
is greatly helpful for explicit model building~\cite{Xing:2015fdg,Xing:2022uax}.

\item     The well-known Jarlskog invariant of leptonic CP
violation~\cite{Jarlskog:1985ht,Wu:1985ea}, which measures
the universal strength of CP-violating effects in neutrino oscillations,
is given as
\begin{eqnarray}
{\cal J}^{}_\nu \equiv {\rm Im}\left(U^{}_{e 2} U^{}_{\mu 3} U^*_{e 3}
U^*_{\mu 2}\right) \simeq \frac{1}{6} \xi \left[\sqrt{2}
- \left(\sqrt{2} + A\right) \xi^2\right] \sin\delta \; . \hspace{0.3cm}
\end{eqnarray}
So ${\cal J}^{}_\nu \simeq 3.5 \times 10^{-2} \sin\delta$ holds in the
leading-order approximation. Given $\delta \sim -\pi/2$, for instance,
the size of the leptonic Jarlskog invariant will be about a thousand
times larger than that of its counterpart in the quark
sector~\cite{ParticleDataGroup:2022pth}.
\end{itemize}
Note that one only needs to keep the leading-order terms of ${\cal A}^{}_1$,
${\cal A}^{}_2$, $\Delta^{}_1$, $\Delta^{}_2$, $\Delta^{}_3$ and
${\cal J}^{}_\nu$ in most cases, as they are analytically simple enough
and numerically accurate enough.

\section{The ordering of $|U^{}_{\alpha i}|$}

Taking account of the $U^{}_{\rm TBM}$-based expansion of the PMNS matrix
$U$ in powers of $\xi \simeq 0.149$ in Eq.~(8), we find that the analytical
approximation
\begin{eqnarray}
U & \simeq &
\left(\begin{matrix}
\displaystyle \frac{2}{\sqrt 6} - \frac{1 - \sqrt{2} A}{\sqrt 6} \xi^2
& \displaystyle \frac{1}{\sqrt 3} - \frac{1 + 2\sqrt{2} A}{2\sqrt{3}} \xi^2
& \xi e^{-{\rm i}\delta}
\cr \vspace{-0.45cm} \cr
\displaystyle -\frac{1}{\sqrt 6} - \frac{1}{\sqrt 3} \xi e^{{\rm i}\delta}
+ \frac{\sqrt{2} A - B}{\sqrt 6} \xi^2
& \displaystyle \frac{1}{\sqrt 3} - \frac{1}{\sqrt 6} \xi e^{{\rm i}\delta}
+ \frac{A + \sqrt{2} B}{\sqrt 6} \xi^2
& \displaystyle \frac{1}{\sqrt 2} - \frac{1 + 2 B}{2 \sqrt{2}} \xi^2
\cr \vspace{-0.45cm} \cr
\displaystyle \frac{1}{\sqrt 6} - \frac{1}{\sqrt 3} \xi e^{{\rm i}\delta}
- \frac{\sqrt{2} A + B}{\sqrt 6} \xi^2
& \hspace{0.2cm} \displaystyle -\frac{1}{\sqrt 3}
- \frac{1}{\sqrt 6} \xi e^{{\rm i}\delta}
- \frac{A - \sqrt{2} B}{\sqrt 6} \xi^2 \hspace{0.2cm}
& \displaystyle \frac{1}{\sqrt 2} - \frac{1 - 2 B}{2 \sqrt{2}} \xi^2
\end{matrix} \right) \hspace{0.5cm}
\end{eqnarray}
is actually good enough to fit current neutrino oscillation data.
Let us examine to what extent one may identify the ordering of
$|U^{}_{\alpha i}|$ (for $\alpha = e, \mu, \tau$ and $i = 1, 2, 3$)
\footnote{Possible ordering of the PMNS moduli has also been discussed
in Ref.~\cite{Denton:2020exu} by doing a careful {\it numerical}
analysis of the available experimental data on neutrino oscillations.
Our present study focuses on the {\it analytical} approximations of
$|U^{}_{\alpha i}|$ which are in most cases insensitive to the
fluctuations of their global-fit values.}.
\begin{itemize}
\item     For the matrix elements in the first row of $U$, it is
easy to identify $|U^{}_{e 1}| > |U^{}_{e 2}| > |U^{}_{e 3}|$.
In fact, $|U^{}_{e 1}|$ and $|U^{}_{e 3}|$ are the largest and smallest
moduli among the nine elements of $U$.

\item     For the matrix elements in the third column of $U$, the sign
of $B$ is crucial as it determines whether $|U^{}_{\mu 3}|$ is larger
or smaller than its counterpart $|U^{}_{\tau 3}|$.
This point is obviously supported by
$|U^{}_{\tau 3}|^2 - |U^{}_{\mu 3}|^2 \simeq 2 B \xi^2$ obtained from
Eq.~(10). We are therefore left with
$|U^{}_{\tau 3}| \geq |U^{}_{\mu 3}| > |U^{}_{e 3}|$ for $B \geq 0$,
or $|U^{}_{\mu 3}| \geq |U^{}_{\tau 3}| > |U^{}_{e 3}|$ for $B \leq 0$.

\item     For the matrix elements in the second and third rows of $U$,
the smallness of $\xi$ assures that $|U^{}_{\mu 1}| < |U^{}_{\mu 2}|
< |U^{}_{\mu 3}|$ and $|U^{}_{\tau 1}| < |U^{}_{\tau 2}|
< |U^{}_{\tau 3}|$ hold. This observation is independent of the values
of $A$, $B$ and $\delta$ in the Wolfenstein-like expansion of $U$
proposed above.

\item     To compare between the magnitudes of $U^{}_{\tau i}$ and
$U^{}_{\mu i}$ (for $i = 1, 2$), we may simplify the expressions of
$\Delta^{}_1$ and $\Delta^{}_2$ in Eq.~(10) as follows:
\begin{eqnarray}
|U^{}_{\tau 1}|^2 - |U^{}_{\mu 1}|^2 & \simeq &
- \frac{2}{3} \xi \left(\sqrt{2} \cos\delta + B \xi \right) \; ,
\nonumber \\
|U^{}_{\tau 2}|^2 - |U^{}_{\mu 2}|^2 & \simeq &
+ \frac{2}{3} \xi \left(\sqrt{2} \cos\delta - 2 B \xi\right) \; .
\hspace{0.5cm}
\end{eqnarray}
It becomes clear that $|U^{}_{\tau 1}| \leq |U^{}_{\mu 1}|$ will
hold if $\sqrt{2} \cos\delta + B \xi \geq 0$ is satisfied; and
$|U^{}_{\tau 1}| \geq |U^{}_{\mu 1}|$ will hold if
$\sqrt{2} \cos\delta + B \xi$ flips its sign. On the other hand,
$|U^{}_{\tau 2}| \geq |U^{}_{\mu 2}|$ will
hold if $\cos\delta \geq \sqrt{2} B \xi$ is satisfied; and
$|U^{}_{\tau 2}| \leq |U^{}_{\mu 2}|$ will hold if
$\cos\delta \leq \sqrt{2} B \xi$ is satisfied.

\item     Whether $|U^{}_{e 2}|$ can be larger or smaller
than $|U^{}_{\mu 2}|$ or $|U^{}_{\tau 2}|$
is another open question before the quadrant of the CP-violating
phase $\delta$ is surely determined. The reason is simply that
\begin{eqnarray}
|U^{}_{e 2}|^2 & \simeq & |U^{}_{\mu 2}|^2 - \frac{1}{2}
\left(1 + 2\sqrt{2} A\right) \xi^2 + \frac{\sqrt 2}{3} \xi
\left(\cos\delta - \sqrt{2} B \xi\right) \;
\nonumber \\
& \simeq & |U^{}_{\tau 2}|^2 - \frac{1}{2}
\left(1 + 2\sqrt{2} A\right) \xi^2 - \frac{\sqrt 2}{3} \xi
\left(\cos\delta - \sqrt{2} B \xi\right) \;  \hspace{0.5cm}
\end{eqnarray}
holds to the accuracy of ${\cal O}(\xi^2)$. In case of
$\cos\delta = 0$, however, the sign of $B$ will play a part.
\end{itemize}
A summary of the above discussions leads us to the following most
likely ordering of the nine matrix elements of $U$ in magnitude:
\begin{eqnarray}
|U^{}_{e 1}| > \{ |U^{}_{\mu 3}| , |U^{}_{\tau 3}| \} >
\{ |U^{}_{e 2}| , |U^{}_{\mu 2}| , |U^{}_{\tau 2}| \} >
\{ |U^{}_{\mu 1}| , |U^{}_{\tau 1}| \} > |U^{}_{e 3}| \; ,
\end{eqnarray}
where the ordering of the {\it bracketed} moduli remains unidentifiable
from the present neutrino oscillation data. That is why the
next-generation long-baseline neutrino oscillation experiments
aim to pin down the octant of $\theta^{}_{23}$ (or equivalently,
the sign of $B$) and the quadrant of $\delta$.

To be more specific, let us simply take the best-fit value of $\delta$
to illustrate the ordering of the nine PMNS moduli $|U^{}_{\alpha i}|$ (for
$\alpha = e, \mu, \tau$ and $i = 1, 2, 3$).
Given~\cite{Gonzalez-Garcia:2021dve,Capozzi:2021fjo}
\begin{eqnarray}
{\rm NMO:} \quad \delta \simeq -0.711\pi \; , \quad A \simeq 1.463 \; , \quad
B \simeq 2.225 \; ,
\end{eqnarray}
where the values of $A$ and $B$ are directly extracted from Eq.~(4), we obtain
$|U^{}_{\tau 1}| > |U^{}_{\mu 1}|$ from Eq.~(13),
$|U^{}_{\tau 2}| < |U^{}_{e 2}| < |U^{}_{\mu 2}|$ from Eq.~(14) and
$|U^{}_{\tau 3}| > |U^{}_{\mu 3}|$ from Eq.~(10). As a result,
\begin{eqnarray}
{\rm NMO:} \quad |U^{}_{e 1}| > |U^{}_{\tau 3}| > |U^{}_{\mu 3}| >
|U^{}_{\mu 2}| > |U^{}_{e 2}| > |U^{}_{\tau 2}| >
|U^{}_{\tau 1}| > |U^{}_{\mu 1}| > |U^{}_{e 3}| \; .
\end{eqnarray}
In the IMO case, we input~\cite{Gonzalez-Garcia:2021dve,Capozzi:2021fjo}
\begin{eqnarray}
{\rm IMO:} \quad \delta \simeq -0.467\pi \; , \quad A \simeq 1.463 \; , \quad
B \simeq -3.135 \; ,
\end{eqnarray}
where the values of $A$ and $B$ are directly extracted from Eq.~(5), and
then find $|U^{}_{\tau 1}| > |U^{}_{\mu 1}|$ and
$|U^{}_{e 2}| < |U^{}_{\mu 2}| < |U^{}_{\tau 2}|$ from Eq.~(14) and
$|U^{}_{\tau 3}| < |U^{}_{\mu 3}|$ from Eq.~(10). As a consequence,
\begin{eqnarray}
{\rm IMO:} \quad |U^{}_{e 1}| > |U^{}_{\mu 3}| > |U^{}_{\tau 3}| >
|U^{}_{\tau 2}| > |U^{}_{\mu 2}| > |U^{}_{e 2}| >
|U^{}_{\tau 1}| > |U^{}_{\mu 1}| > |U^{}_{e 3}| \; .
\end{eqnarray}
We see that the nine PMNS matrix elements do not have a clearly layered hierarchy
in magnitude, as compared with the four-layered ordering of the nine CKM moduli
shown in Eq.~(1). The reason behind this difference should be closely related
to the underlying mechanism responsible for the origin of tiny neutrino masses,
although it remains vague and unclear at present.

\section{The unitarity triangle}

The ``appearance" neutrino oscillation $\nu^{}_\mu \to \nu^{}_e$ and
its CP-conjugated process $\overline{\nu}^{}_\mu \to \overline{\nu}^{}_e$
are the only realistic channels to measure leptonic CP violation in
a long-baseline oscillation experiment like T2K~\cite{T2K:2023smv}.
It is the so-called unitarity triangle $\triangle^{}_\tau$~\cite{Fritzsch:1999ee}
defined by the orthogonality relation
$U^{}_{e 1} U^*_{\mu 1} + U^{}_{e 2} U^*_{\mu 2} + U^{}_{e 3} U^*_{\mu 3} = 0$
in the complex plane that is directly related to $\nu^{}_\mu \to \nu^{}_e$ and
$\overline{\nu}^{}_\mu \to \overline{\nu}^{}_e$ oscillations. In view
of the PDG-advocated phase convention of $U$ taken in Eq.~(6) or (8), we find that
it is more convenient to use the side $U^{}_{e 3} U^*_{\mu 3}$ to
rescale the three sides of $\triangle^{}_\tau$~\cite{Xing:2019tsn}.
In this case, we simply arrive at
\begin{eqnarray}
\triangle^{\prime}_\tau:
\quad \frac{U^{}_{e 1} U^*_{\mu 1}}{U^{}_{e 3} U^*_{\mu 3}}
+ \frac{U^{}_{e 2} U^*_{\mu 2}}{U^{}_{e 3} U^*_{\mu 3}} + 1 = 0 \; ,
\end{eqnarray}
where the two sloping sides are given by
\begin{eqnarray}
\frac{U^{}_{e 1} U^*_{\mu 1}}{U^{}_{e 3} U^*_{\mu 3}}
& \simeq & -\frac{\sqrt 2}{3 \xi} e^{{\rm i}\delta} - \frac{2}{3} \; ,
\nonumber \\
\frac{U^{}_{e 2} U^*_{\mu 2}}{U^{}_{e 3} U^*_{\mu 3}}
& \simeq & +\frac{\sqrt 2}{3 \xi} e^{{\rm i}\delta} - \frac{1}{3} \; ,
\hspace{0.5cm}
\end{eqnarray}
to a good degree of accuracy. Just taking $\delta \simeq -\pi/2$ and
$\xi \simeq 0.149$ for example, we obtain the numerical results
$|U^{}_{e 1} U^*_{\mu 1}|/|U^{}_{e 3} U^*_{\mu 3}| \simeq 3.23$
and $|U^{}_{e 2} U^*_{\mu 2}|/|U^{}_{e 3} U^*_{\mu 3}| \simeq 3.18$.
Namely, two of the three sides of $\triangle^{\prime}_\tau$
are about three times longer than the shortest one in magnitude, as
illustrated by the solid black triangle in Fig.~1.
Note that the height of $\triangle^{\prime}_\tau$, denoted as
${\cal J}^\prime_\nu$, is correlated with the Jarlskog invariant
${\cal J}^{}_\nu$ as follows:
\begin{eqnarray}
{\cal J}^\prime_\nu = \frac{{\cal J}^{}_\nu}{|U^{}_{e 3} U^*_{\mu 3}|^2}
\simeq \frac{1}{3\xi} \left[\sqrt{2} - \left(A - 2\sqrt{2} B\right)
\xi^2 \right] \sin\delta \; ,
\end{eqnarray}
where the expression of ${\cal J}^{}_\nu$ obtained in Eq.~(11) has been
used. We are then left with the result
${\cal J}^\prime_\nu \simeq \sqrt{2}\sin\delta/
\left(3\xi\right)$ in the leading-order approximation, as clearly
indicated by the imaginary parts of the two sloping sides of
$\triangle^{\prime}_\tau$ in Eq.~(21).
\begin{figure}[t]
\centering
\includegraphics[width = 0.65\linewidth]{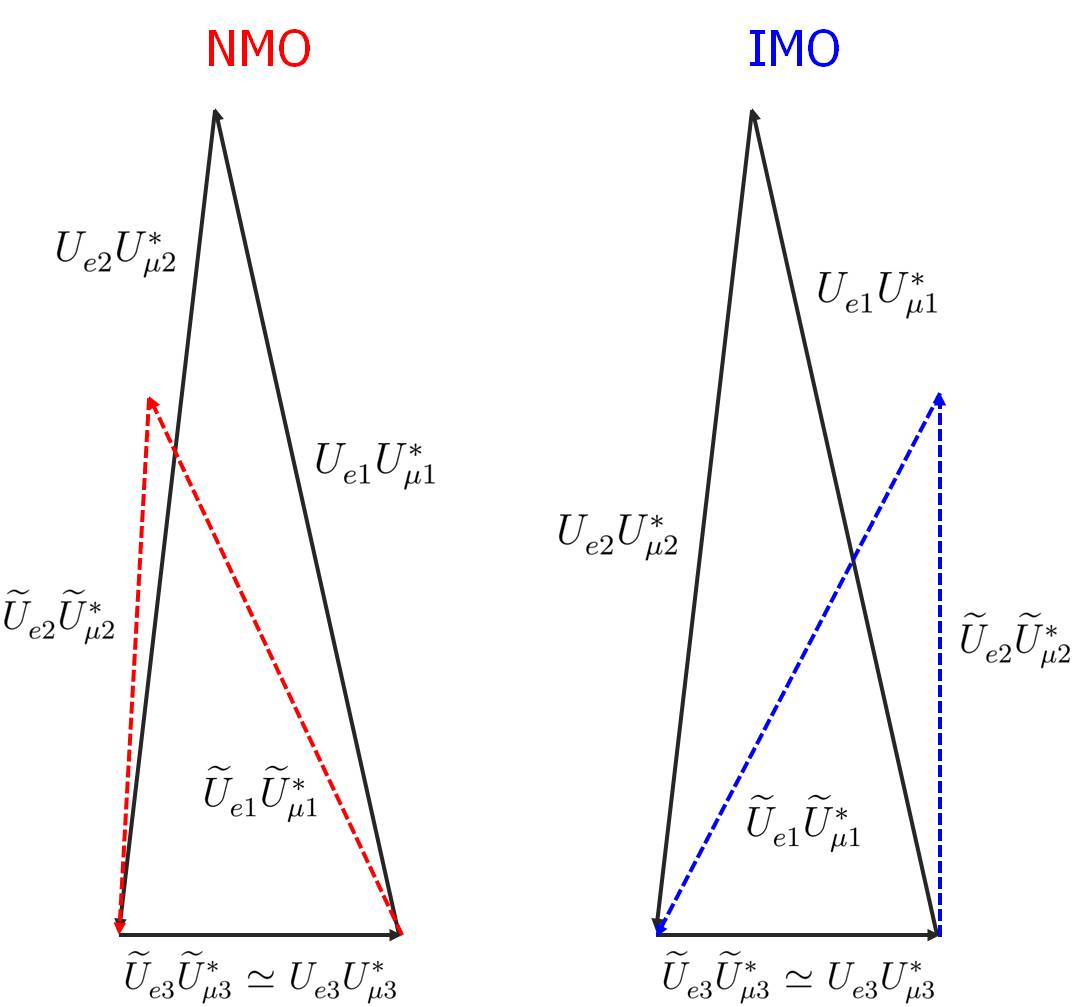}
\vspace{-0.2cm}
\caption{An illustration of the geometrical shapes of the rescaled unitarity
triangle $\triangle^\prime_\tau$ (solid black) and its effective
matter-corrected counterpart $\widetilde{\triangle}^\prime_\tau$
(dashed red for the NMO case and dashed blue for the IMO case)
in the complex plane, where $\delta \simeq -\pi/2$ has typically been input.}
\end{figure}

It is well known that the terrestrial matter effects on $\nu^{}_\mu \to \nu^{}_e$
and $\overline{\nu}^{}_\mu \to \overline{\nu}^{}_e$ oscillations are
not very significant in the T2K and Hyper-Kamiokande experiments with
the baseline length $L \simeq 295 ~{\rm km}$ and the typical beam
energy $E \simeq 0.6 ~{\rm GeV}$~\cite{ParticleDataGroup:2022pth},
but such effects can modify the shape of the above unitarity triangle
and thus modify the Jarlskog invariant of CP violation to some extent.
To see this point more clearly, let us consider the {\it effective}
rescaled unitarity triangle
\begin{eqnarray}
\widetilde{\triangle}^{\prime}_\tau:
\quad \frac{\widetilde{U}^{}_{e 1} \widetilde{U}^*_{\mu 1}}
{\widetilde{U}^{}_{e 3} \widetilde{U}^*_{\mu 3}}
+ \frac{\widetilde{U}^{}_{e 2} \widetilde{U}^*_{\mu 2}}
{\widetilde{U}^{}_{e 3} \widetilde{U}^*_{\mu 3}} + 1 = 0 \; ,
\end{eqnarray}
where $\widetilde{U}^{}_{e i}$ and $\widetilde{U}^{}_{\mu i}$ (for
$i = 1, 2, 3$) denote the effective PMNS matrix elements in matter.
Following the analytical approximations made in Ref.~\cite{Xing:2016ymg},
we obtain $\widetilde{U}^{}_{e 3} \widetilde{U}^*_{\mu 3}
\simeq U^{}_{e 3} U^*_{\mu 3}$ and
\begin{eqnarray}
\frac{\widetilde{U}^{}_{e 1} \widetilde{U}^*_{\mu 1}}
{\widetilde{U}^{}_{e 3} \widetilde{U}^*_{\mu 3}}
& \simeq & \frac{\alpha}{\epsilon} \cdot
\frac{U^{}_{e 1} U^*_{\mu 1}}{U^{}_{e 3} U^*_{\mu 3}}
- \frac{\epsilon - \alpha - \beta}{2\epsilon} \; ,
\nonumber \\
\frac{\widetilde{U}^{}_{e 2} \widetilde{U}^*_{\mu 2}}
{\widetilde{U}^{}_{e 3} \widetilde{U}^*_{\mu 3}}
& \simeq & \frac{\alpha}{\epsilon} \cdot
\frac{U^{}_{e 2} U^*_{\mu 2}}{U^{}_{e 3} U^*_{\mu 3}}
- \frac{\epsilon - \alpha + \beta}{2\epsilon} \; ,
\hspace{0.5cm}
\end{eqnarray}
to a good degree of accuracy, where
$\alpha \equiv \Delta m^2_{21}/\Delta m^2_{31}$,
$\beta \equiv \mathbb{A}/\Delta m^2_{31}$,
$\mathbb{A} \equiv 2 \sqrt{2} \hspace{0.05cm}
G^{}_{\rm F} N^{}_e E$, and
\begin{eqnarray}
\epsilon \equiv \sqrt{\alpha^2 - 2\left(|U^{}_{e 1}|^2 -
|U^{}_{e 2}|^2\right) \alpha\beta + \left(1 - |U^{}_{e 3}|^2
\right)^2 \beta^2} \;
\end{eqnarray}
with $\Delta m^2_{ij}$ (for $i, j = 1, 2, 3$) being the neutrino
mass-squared differences, $G^{}_{\rm F}$ being the Fermi constant,
$N^{}_e$ being the terrestrial background density of electrons,
and $E$ being the neutrino beam
energy~\cite{Wolfenstein:1977ue,Mikheyev:1985zog}. Substituting
Eq.~(21) into Eq.~(24), we arrive at
\begin{eqnarray}
\frac{\widetilde{U}^{}_{e 1} \widetilde{U}^*_{\mu 1}}
{\widetilde{U}^{}_{e 3} \widetilde{U}^*_{\mu 3}}
& \simeq & -\left(\frac{\sqrt 2}{3 \xi} e^{{\rm i}\delta} + \frac{1}{6}\right)
\frac{\alpha}{\epsilon} - \frac{1}{2} \left(1 - \frac{\beta}{\epsilon}\right) \; ,
\nonumber \\
\frac{\widetilde{U}^{}_{e 2} \widetilde{U}^*_{\mu 2}}
{\widetilde{U}^{}_{e 3} \widetilde{U}^*_{\mu 3}}
& \simeq & +\left(\frac{\sqrt 2}{3 \xi} e^{{\rm i}\delta} + \frac{1}{6}\right)
\frac{\alpha}{\epsilon} - \frac{1}{2} \left(1 + \frac{\beta}{\epsilon}\right) \; .
\hspace{0.5cm}
\end{eqnarray}
This result shows that the terrestrial matter effects can obviously
modify the rescaled unitarity triangle $\triangle^\prime_\tau$ in
vacuum. Taking account of
$\widetilde{\cal J}^{}_\nu/{\cal J}^{}_\nu \simeq |\alpha|/\epsilon$
obtained in Ref.~\cite{Xing:2016ymg}, where $\widetilde{\cal J}^{}_\nu$
denotes the effective Jarlskog invariant in matter, we may similarly
achieve the height of $\widetilde{\triangle}^\prime_\tau$:
\begin{eqnarray}
\widetilde{\cal J}^\prime_\nu = \frac{\widetilde{\cal J}^{}_\nu}
{\big|\widetilde{U}^{}_{e 3} \widetilde{U}^*_{\mu 3}\big|^2}
\simeq \frac{\alpha}{3 \epsilon \xi}
\left[\sqrt{2} - \left(A - 2\sqrt{2} B\right) \xi^2 \right] \sin\delta \; ,
\end{eqnarray}
whose leading term is certainly consistent with the imaginary parts
of the two sloping sides of $\widetilde{\triangle}^\prime_\tau$
that can directly be seen from Eq.~(26).

To illustrate, one may typically
take $\mathbb{A} \simeq 2.28 \times 10^{-4} ~{\rm eV}^2 \left(E/{\rm GeV}
\right)$ for a neutrino trajectory through the Earth's
crust~\cite{Mocioiu:2000st}, which is suitable for the realistic
ongoing and upcoming long-baseline neutrino oscillation
experiments. Of course, the matter parameter $\mathbb{A}$ should
flip its sign for an antineutrino beam, and both $\alpha$ and
$\beta$ are sensitive to the neutrino mass ordering (i.e., the sign
of $\Delta m^2_{31}$). Given the best-fit values
$\Delta m^2_{21} \simeq 7.41 \times 10^{-5} ~{\rm eV}^2$,
$\Delta m^2_{31} \simeq 2.51 \times 10^{-3} ~{\rm eV}^2$ (NMO) or
$\Delta m^2_{31} \simeq -2.41 \times 10^{-3} ~{\rm eV}^2$
(IMO)~\cite{Gonzalez-Garcia:2021dve,Capozzi:2021fjo}, we have
$\alpha \simeq 2.95 \times 10^{-2}$ (NMO) or
$\alpha \simeq -3.07 \times 10^{-2}$ (IMO), together with
$\beta \simeq 5.45 \times 10^{-2}$ (NMO) or
$\beta \simeq -5.68 \times 10^{-2}$ (IMO) for the T2K and
Hyper-Kamiokande neutrino oscillation experiments with
$E \simeq 0.6 ~{\rm GeV}$. In this case we obtain
$\epsilon \simeq 4.97 \times 10^{-2}$ (NMO) or
$\epsilon \simeq 5.18 \times 10^{-2}$ (IMO) after taking
into account the best-fit values of $\theta^{}_{12}$ and
$\theta^{}_{13}$ given in Eqs.~(2) and (3). Fig.~1 illustrates
how the geometrical shape of $\triangle^\prime_\tau$ changes
as a consequence of the terrestrial matter effects on neutrino
oscillations in the NMO and IMO cases. It becomes clear that the
area of $\widetilde{\triangle}^\prime_\tau$ is remarkably
smaller than that of $\triangle^\prime_\tau$, and their
ratio is simply governed by
$\widetilde{\cal J}^{}_\nu/{\cal J}^{}_\nu \simeq |\alpha|/\epsilon$
as discussed above.

It is worth pointing out that $\theta^{}_{13}$, $\theta^{}_{23}$
and $\delta$ are essentially insensitive to terrestrial matter
effects in the $E \lesssim 1 ~{\rm GeV}$ region~\cite{Minakata:2000ee,
Akhmedov:2001kd,Xing:2003ez}, and
$\sin 2\widetilde{\theta}^{}_{12}/\sin 2\theta^{}_{12}
\simeq \widetilde{\cal J}^{}_\nu/{\cal J}^{}_\nu \simeq
|\alpha|/\epsilon$ holds as a good approximation~\cite{Xing:2016ymg}.
These observations imply that $\xi$, $B$ and $\delta$ in our
Wolfenstein-like expansion of the PMNS matrix $U$ are also expected
to be insensitive to terrestrial matter effects in an
accelerator-based neutrino oscillation experiment with
$E \lesssim 1 ~{\rm GeV}$, and only $A$ is an exception. To be
specific, we take $\widetilde{\theta}^{}_{12}
\simeq \theta^{(0)}_{12} - \widetilde{A} \xi^2$ and then obtain
\begin{eqnarray}
\widetilde{A} \xi^2 \simeq A \xi^2 \frac{|\alpha|}{\epsilon}
+ \sqrt{2} \left(1 - \frac{|\alpha|}{\epsilon}\right) \; .
\end{eqnarray}
The second term on the right-hand side of Eq.~(28) is in general
unsuppressed as it arises from $\theta^{(0)}_{12} \simeq 35.26^\circ$.
Although $\widetilde{A}$ itself may be far above
${\cal O}(1)$, it does not really point to a significant matter
effect simply because it is not directly related to any observable
of neutrino oscillations.

Note that the terrestrial matter effects are expected to be far more
significant in the upcoming DUNE neutrino oscillation experiment with
the baseline length $L \simeq 1300 ~{\rm km}$ and the beam energy
$E \in [2, 3]~{\rm GeV}$~\cite{DUNE:2015lol}. In this case one may
instead use Freund's analytical approximations for the neutrino oscillation
parameters in matter~\cite{Freund:2001pn} to discuss the matter-modified
PMNS lepton flavor mixing matrix and the corresponding unitarity triangles.

\section{Further discussions}

Motivated by the fact that the fine structure of quark flavor mixing
can well be understood in the Wolfenstein expansion of the CKM matrix
$V$, we have proposed a similar expansion of the PMNS matrix $U$ in the
basis of the tri-bimaximal mixing pattern $U^{}_{\rm TMB}$ towards understanding
the fine structure of lepton flavor mixing. The corresponding expansion
parameters are $\lambda \simeq 0.225$ for $V$ and $\xi \simeq 0.149$ for $U$,
and thus we can directly arrive at
\begin{eqnarray}
\lim_{\lambda \to 0} V = I \; , \quad
\lim_{\xi \to 0} U = U^{}_{\rm TBM} \; ,
\end{eqnarray}
in the limits of vanishing or vanishingly small $\lambda$ and $\xi$.
While $I$ is unique in the quark sector, $U^{}_{\rm TBM}$ is just our
choice in the lepton sector. However, we have argued that it is rather
reasonable to choose $U^{}_{\rm TBM}$ as an expansion basis of
$U$ since this constant pattern is particularly favored
from the point of view of model building with the help of an underlying
discrete flavor symmetry.

It makes sense to ask what new information can be achieved from the proposed
Wolfenstein-like expansion of the $U^{}_{\rm TBM}$-based PMNS matrix $U$ as
compared with the exact PDG-advocated parametrization of $U$ in terms of
the Euler-like angles and phases. We find that our Wolfenstein-like
parametrization {\it does} have some phenomenological merits.
\begin{itemize}
\item     It can easily reflect the relative magnitudes of the nine
matrix elements of $U$ to a reasonable degree of accuracy.
For example, the ratios $|U^{}_{e 1}| : |U^{}_{\mu 1}| : |U^{}_{\tau 1}| \simeq
2: 1: 1$ and $|U^{}_{e 3}| : |U^{}_{\mu 3}| : |U^{}_{\tau 3}| \simeq
\sqrt{2}\hspace{0.05cm} \xi : 1 : 1$ are the straightforward consequences
of Eq.~(12) in no need of doing any algebraic calculations, and they are
essentially consistent with the numerical results obtained from a global
fit of current neutrino oscillation data~\cite{Gonzalez-Garcia:2021dve,Capozzi:2021fjo}.
Such an advantage is analogous to the fact that $|V^{}_{ub}| : |V^{}_{cb}| : |V^{}_{us}|
: |V^{}_{ud}| \simeq \lambda^3 : \lambda^2 : \lambda : 1$ can be directly
read off from the Wolfenstein parametrization of the CKM matrix $V$.

\item     It can help establish some simple and testable relations between different
PMNS matrix elements which are associated with different neutrino oscillation
experiments. For instance, the relation $|U^{}_{\mu 2}|^2 + |U^{}_{\tau 2}|^2
\simeq 2 |U^{}_{e 2}|^2$ holds up to a small correction of ${\cal O}(\xi^2)$, as
can be seen from Eq.~(14); and ${\cal J}^{}_\nu \simeq |U^{}_{e 1}|
|U^{}_{e 2}| |U^{}_{e 3}| \sin\delta$ holds to the same degree of accuracy,
as can be easily derived from Eqs.~(11) and (12). In comparison, it is difficult
to obtain such instructive relations from the exact Euler-like parametrization
in Eq.~(6).

\item     It can largely simplify all the leading-order calculations and
some of the next-to-leading-order calculations of the observable effects
regarding lepton flavor mixing and CP violation. A typical example of this
kind --- the leptonic unitarity triangle $\triangle^{}_\tau$ has been presented
in section 4, from which one can see that even the analysis of terrestrial matter
effects is simplified to some extent in our Wolfenstein-like parametrization of $U$.
The reason for this simplification is of course that $U$ has been expanded in the 
basis of the constant flavor mixing pattern $U^{}_{\rm TBM}$ and in powers of the 
well-measured small parameter $\xi \equiv |U^{}_{e3}|$.
\end{itemize}
So our approach proves to be useful, and more of its applications
in neutrino phenomenology (e.g., on the flavor distribution of ultrahigh-energy
cosmic neutrinos) will be explored elsewhere.

Finally, it is worth pointing out that the small expansion parameters like $\lambda$
and $\xi$ usually play a role of characterizing possible flavor symmetry breaking 
effects or quantum corrections in building an explicit top-down model of fermion 
mass generation and flavor mixing at a proper energy scale. In this case the CKM 
matrix $V$ and the PMNS matrix $U$ are very likely to satisfy the intriguing 
limits like Eq.~(29). Although $U^{}_{\rm TBM}$ itself is not a unique choice along 
this line of thought, the Wolfenstein-like expansions of $V$ and $U$ should make 
sense in any case.

\section*{Acknowledgements}

I am indebted to Jihong Huang for useful discussions. This work was supported
by the National Natural Science Foundation of China
under grant No. 12075254 and grant No. 11835013.



\begin{thebibliography}{99}
\addtolength{\itemsep}{-1.5ex}

\bibitem{Cabibbo:1963yz}
  N.~Cabibbo,
  Unitary Symmetry and Leptonic Decays,
  Phys.\ Rev.\ Lett.\  {\bf 10} (1963) 531.

\bibitem{Kobayashi:1973fv}
  M.~Kobayashi and T.~Maskawa,
  CP Violation in the Renormalizable Theory of Weak Interaction,
  Prog.\ Theor.\ Phys.\  {\bf 49} (1973) 652.

\bibitem{ParticleDataGroup:2022pth}
R.~L.~Workman \textit{et al.} [Particle Data Group],
``Review of Particle Physics,''
PTEP \textbf{2022} (2022), 083C01

\bibitem{Wolfenstein:1983yz}
  L.~Wolfenstein,
  Parametrization of the Kobayashi-Maskawa Matrix,
  Phys.\ Rev.\ Lett.\  {\bf 51} (1983) 1945.

\bibitem{Xing:1996it}
Z.~Z.~Xing,
``On the hierarchy of quark mixings,''
Nuovo Cim. A \textbf{109} (1996), 115-118

\bibitem{Xing:2020ijf}
Z.~z.~Xing,
``Flavor structures of charged fermions and massive neutrinos,''
Phys. Rept. \textbf{854} (2020), 1-147
[arXiv:1909.09610 [hep-ph]].

\bibitem{Xing:2002az}
  Z.~z.~Xing,
  Wolfenstein - like parametrization of the neutrino mixing matrix,
  J.\ Phys.\ G {\bf 29} (2003) 2227
  [hep-ph/0211465].

\bibitem{Kaus:2002zm}
P.~Kaus and S.~Meshkov,
``Neutrino mass matrix and hierarchy,''
AIP Conf. Proc. \textbf{672} (2003) no.1, 117-125
[arXiv:hep-ph/0211338 [hep-ph]].

\bibitem{Gupta:2013vva}
V.~Gupta, G.~S\'anchez-Col\'on, S.~Rajpoot and H.~C.~Wang,
``Lepton flavor mixing in the Wolfenstein scheme,''
Phys. Rev. D \textbf{87} (2013), 073009
[arXiv:1304.1065 [hep-ph]].

\bibitem{Fritzsch:1995dj}
H.~Fritzsch and Z.~Z.~Xing,
``Lepton mass hierarchy and neutrino oscillations,''
Phys. Lett. B \textbf{372} (1996), 265-270
[arXiv:hep-ph/9509389 [hep-ph]].

\bibitem{Fritzsch:1998xs}
H.~Fritzsch and Z.~z.~Xing,
``Large leptonic flavor mixing and the mass spectrum of leptons,''
Phys. Lett. B \textbf{440} (1998), 313-318
[arXiv:hep-ph/9808272 [hep-ph]].

\bibitem{Feruglio:2019ybq}
F.~Feruglio and A.~Romanino,
``Lepton flavor symmetries,''
Rev. Mod. Phys. \textbf{93} (2021) no.1, 015007
[arXiv:1912.06028 [hep-ph]].

\bibitem{King:2017guk}
S.~F.~King,
``Unified Models of Neutrinos, Flavour and CP Violation,''
Prog. Part. Nucl. Phys. \textbf{94} (2017), 217-256
[arXiv:1701.04413 [hep-ph]].

\bibitem{Ding:2024ozt}
G.~J.~Ding and J.~W.~F.~Valle,
``The symmetry approach to quark and lepton masses and mixing,''
[arXiv:2402.16963 [hep-ph]].

\bibitem{Harrison:2002er}
P.~F.~Harrison, D.~H.~Perkins and W.~G.~Scott,
``Tri-bimaximal mixing and the neutrino oscillation data,''
Phys. Lett. B \textbf{530} (2002), 167
[arXiv:hep-ph/0202074 [hep-ph]].

\bibitem{Xing:2002sw}
Z.~z.~Xing,
``Nearly tri-bimaximal neutrino mixing and CP violation,''
Phys. Lett. B \textbf{533} (2002), 85-93
[arXiv:hep-ph/0204049 [hep-ph]].

\bibitem{He:2003rm}
X.~G.~He and A.~Zee,
``Some simple mixing and mass matrices for neutrinos,''
Phys. Lett. B \textbf{560} (2003), 87-90
[arXiv:hep-ph/0301092 [hep-ph]].

\bibitem{Li:2004dn}
N.~Li and B.~Q.~Ma,
``Parametrization of neutrino mixing matrix in tri-bimaximal mixing pattern,''
Phys. Rev. D \textbf{71} (2005), 017302
[arXiv:hep-ph/0412126 [hep-ph]].

\bibitem{King:2007pr}
S.~F.~King,
``Parametrizing the lepton mixing matrix in terms of deviations from tri-bimaximal mixing,''
Phys. Lett. B \textbf{659} (2008), 244-251
[arXiv:0710.0530 [hep-ph]].

\bibitem{King:2009qt}
S.~F.~King,
``Tri-bimaximal Neutrino Mixing and $\theta_{13}$,''
Phys. Lett. B \textbf{675} (2009), 347-351
[arXiv:0903.3199 [hep-ph]].

\bibitem{King:2012vj}
S.~F.~King,
``Tri-bimaximal-Cabibbo Mixing,''
Phys. Lett. B \textbf{718} (2012), 136-142
[arXiv:1205.0506 [hep-ph]].

\bibitem{Liu:2014gla}
Z.~Liu and Y.~L.~Wu,
``Leptonic CP Violation and Wolfenstein Parametrization for Lepton Mixing,''
Phys. Lett. B \textbf{733} (2014), 226-232
[arXiv:1403.2440 [hep-ph]].

\bibitem{Gonzalez-Garcia:2021dve}
M.~C.~Gonzalez-Garcia, M.~Maltoni and T.~Schwetz,
``NuFIT: Three-Flavour Global Analyses of Neutrino Oscillation Experiments,''
Universe \textbf{7} (2021) no.12, 459
[arXiv:2111.03086 [hep-ph]];
NuFit webpage, http://www.nu-fit.org.

\bibitem{Capozzi:2021fjo}
F.~Capozzi, E.~Di Valentino, E.~Lisi, A.~Marrone, A.~Melchiorri and A.~Palazzo,
``Unfinished fabric of the three neutrino paradigm,''
Phys. Rev. D \textbf{104} (2021) no.8, 083031
[arXiv:2107.00532 [hep-ph]].

\bibitem{Giunti:2002ye}
C.~Giunti and M.~Tanimoto,
``Deviation of neutrino mixing from bimaximal,''
Phys. Rev. D \textbf{66} (2002), 053013
[arXiv:hep-ph/0207096 [hep-ph]].

\bibitem{DayaBay:2018yms}
D.~Adey \textit{et al.} [Daya Bay],
``Measurement of the Electron Antineutrino Oscillation with 1958 Days of Operation at Daya Bay,''
Phys. Rev. Lett. \textbf{121} (2018) no.24, 241805
[arXiv:1809.02261 [hep-ex]].

\bibitem{Xing:2015fdg}
Z.~z.~Xing and Z.~h.~Zhao,
``A review of $\mu$-$\tau$ flavor symmetry in neutrino physics,''
Rept. Prog. Phys. \textbf{79} (2016) no.7, 076201
[arXiv:1512.04207 [hep-ph]].

\bibitem{Xing:2022uax}
Z.~z.~Xing,
``The $\mu$-$\tau$ reflection symmetry of Majorana neutrinos,''
Rept. Prog. Phys. \textbf{86} (2023) no.7, 076201
[arXiv:2210.11922 [hep-ph]].

\bibitem{Xing:2002sx}
Z.~z.~Xing,
``Can the lepton flavor mixing matrix be symmetric?,''
Phys. Rev. D \textbf{65} (2002), 113010
[arXiv:hep-ph/0204050 [hep-ph]].

\bibitem{Harrison:2002et}
P.~F.~Harrison and W.~G.~Scott,
``$\mu$-$\tau$ reflection symmetry in lepton mixing and neutrino oscillations,''
Phys. Lett. B \textbf{547} (2002), 219-228
[arXiv:hep-ph/0210197 [hep-ph]].

\bibitem{T2K:2023smv}
K.~Abe \textit{et al.} [T2K],
``Measurements of neutrino oscillation parameters from the T2K experiment using $3.6\times 10^{21}$ protons on target,''
Eur. Phys. J. C \textbf{83} (2023) no.9, 782
[arXiv:2303.03222 [hep-ex]].

\bibitem{Jarlskog:1985ht}
C.~Jarlskog,
``Commutator of the Quark Mass Matrices in the Standard Electroweak Model and a Measure of Maximal CP Nonconservation,''
Phys. Rev. Lett. \textbf{55} (1985), 1039

\bibitem{Wu:1985ea}
D.~d.~Wu,
``The Rephasing Invariants and CP,''
Phys. Rev. D \textbf{33} (1986), 860

\bibitem{Denton:2020exu}
P.~B.~Denton,
``A Return To Neutrino Normalcy,''
[arXiv:2003.04319 [hep-ph]].

\bibitem{Fritzsch:1999ee}
  H.~Fritzsch and Z.~z.~Xing,
  Mass and flavor mixing schemes of quarks and leptons,
  Prog.\ Part.\ Nucl.\ Phys.\  {\bf 45} (2000) 1
  [hep-ph/9912358].

\bibitem{Xing:2019tsn}
Z.~Z.~Xing and D.~Zhang,
``Distinguishing between the twin $b$-flavored unitarity triangles on a circular arc,''
Phys. Lett. B \textbf{803} (2020), 135302
[arXiv:1911.03292 [hep-ph]].

\bibitem{Xing:2016ymg}
Z.~z.~Xing and J.~y.~Zhu,
``Analytical approximations for matter effects on CP violation in the accelerator-based neutrino oscillations with E $\lesssim$ 1 GeV,''
JHEP \textbf{07} (2016), 011
[arXiv:1603.02002 [hep-ph]].

\bibitem{Wolfenstein:1977ue}
L.~Wolfenstein,
``Neutrino Oscillations in Matter,''
Phys. Rev. D \textbf{17} (1978), 2369-2374

\bibitem{Mikheyev:1985zog}
S.~P.~Mikheyev and A.~Y.~Smirnov,
``Resonance Amplification of Oscillations in Matter and Spectroscopy of Solar Neutrinos,''
Sov. J. Nucl. Phys. \textbf{42} (1985), 913-917

\bibitem{Mocioiu:2000st}
I.~Mocioiu and R.~Shrock,
``Matter effects on neutrino oscillations in long baseline experiments,''
Phys. Rev. D \textbf{62} (2000), 053017
[arXiv:hep-ph/0002149 [hep-ph]].

\bibitem{Minakata:2000ee}
H.~Minakata and H.~Nunokawa,
``Measuring leptonic CP violation by low-energy neutrino oscillation experiments,''
Phys. Lett. B \textbf{495} (2000), 369-377
[arXiv:hep-ph/0004114 [hep-ph]].

\bibitem{Akhmedov:2001kd}
E.~K.~Akhmedov, P.~Huber, M.~Lindner and T.~Ohlsson,
``T violation in neutrino oscillations in matter,''
Nucl. Phys. B \textbf{608} (2001), 394-422
[arXiv:hep-ph/0105029 [hep-ph]].

\bibitem{Xing:2003ez}
Z.~z.~Xing,
``Flavor mixing and CP violation of massive neutrinos,''
Int. J. Mod. Phys. A \textbf{19} (2004), 1-80
[arXiv:hep-ph/0307359 [hep-ph]].

\bibitem{DUNE:2015lol}
R.~Acciarri \textit{et al.} [DUNE],
``Long-Baseline Neutrino Facility (LBNF) and Deep Underground Neutrino Experiment (DUNE): Conceptual Design Report, Volume 2: The Physics Program for DUNE at LBNF,''
[arXiv:1512.06148 [physics.ins-det]].

\bibitem{Freund:2001pn}
M.~Freund,
``Analytic approximations for three neutrino oscillation parameters and probabilities in matter,''
Phys. Rev. D \textbf{64} (2001), 053003
[arXiv:hep-ph/0103300 [hep-ph]].

\end{thebibliography}
\end{document}